\documentstyle[preprint,aps]{revtex}

\begin{document}
\draft
\preprint{UAHEP955}
\title{$B$ Decay into Light Gluinos}
\author{Peter Povinec, B. Fenyi, and L. Clavelli}
\address{Dept. of Physics and Astronomy, University of Alabama,
Tuscaloosa, AL 35487, USA}
\date{\today}
\maketitle
\begin{abstract}
Flavor changing interactions of the gluino allow the $b$ quark to
decay
into the strange quark plus a gluino pair if the gluino is in the
ultra
low mass window below 1 GeV. In this case the enhancement of the
nonleptonic $b$ decay could explain the anomalous semileptonic
branching
ratio.
\end{abstract}
\pacs{14.80.Ly, 13.25.+m, 11.30.Pb}

\narrowtext

In the last few years it has been noted by many authors that a light
gluino would
help to explain several anomalies at the $Z$ scale and other
discrepancies between
experiment and theory \cite{uahep953,ccfhpy,clavelli}. Also, the
impact of the light gluino
on the branching ratio $b\rightarrow s \gamma$ has been recently
investigated in
\cite{bhara}. Surprisingly enough, the region of gluino mass below
0.7 GeV
is poorly constrained by experiment \cite{farrar}.
Here, assuming that the gluino is in this low mass window we propose
that the
decay $b\rightarrow s \widetilde g \widetilde g$
might contribute considerably to the $b$ total width thus reducing
the theoretical
prediction of the $B$ semileptonic
branching ratio.

Currently, the experimental value for the semileptonic branching
ratio is
$BR_{SL}(B) \Bigl | _{exp} =(10.43 \pm 0.24)\%$ \cite{pdg},
while the theoretical prediction gives a lower bound of 12.5\%
\cite{bigi}. The last number
includes also perturbative QCD corrections. Non-perturbative QCD
corrections
are  not expected to increase the inclusive non-leptonic widths of
$B$ mesons
significantly. Therefore, the discrepancy between theory and
experiment,
which is at least 14\%, still has to be explained.

One solution proposed in \cite{bigi,kagan} could be a
$\Gamma(b\rightarrow c \bar c s)$
enhancement due to larger than expected non-perturbative corrections
in
the $b\rightarrow c \bar c s$ channel. This would, however, enhance
also the charm multiplicity
to about 1.3 which would be more than 15\% higher than the value
from current
experimental data. Unless the future measurements of the charm
multiplicity
lead to its expected value, it could be plausible that there
are new contributions to charmless $b$ decays, that were unaccounted
for in the
theoretical prediction of $BR_{SL}(B)$. Some authors \cite{falk}
prefer
rather conservative explanations of the $B$ semileptonic branching
ratio
puzzle and suggest that the most natural solution lies within QCD
itself. Nevertheless, they do not exclude the possibility of
scenarios from beyond
the standard model contributing to the solution of the problem.

In the following we show that the $b$ decay to the $s$ quark and a
light gluino pair can easily increase the
nonleptonic branching ratio by 20\% in certain regions of its
parameter space. We also show that this process contributes to the
total width of $b$
by a considerably larger amount than the process $b\rightarrow
s g$ if the gluino is below 1 GeV.

The process we are dealing with is a tree-level
flavor-changing-neutral-current
process with a down type squark in the intermediate state. The decay
rate is
calculated using the quark-squark-gluino Lagrangian \cite{bhara},
given by

\begin{equation}
{\cal L}_{q{\tilde q}{\tilde g}} = i \sqrt{2} g_s {\widetilde
q}_i^{\dagger a} {\overline {\widetilde g}}_\alpha
(\lambda_\alpha/2)_{ab} \left[\Gamma_L^{ip} \frac{1-\gamma_5}{2}
 + \Gamma_R^{ip} \frac{1+\gamma_5}{2} \right] q_p^b,
\label{lag1}
\end{equation}
where $p$ stands for the quark generation (in our case $p=b
\text{ or } s$)
and $i$ labels the squark
states ($i=b_L, b_R, s_L, s_R, d_L, d_R$). The $\lambda_\alpha$
are the eight generators of color SU(3).
The matrices $\Gamma_L$ and $\Gamma_R$ are $(6\times 3)$ matrices
given by

\begin{equation}
\Gamma_L = \widetilde U^\dagger
\pmatrix {~I \cr~0},~~\Gamma_R = \widetilde U^\dagger \pmatrix
{~0 \cr~I };
\end{equation}
where $\widetilde U$ is the matrix that diagonalizes the down-type
squark
mass matrix squared, $M_{\tilde d}^2$. Adopting the notation
of \cite{bhara}
, $M_{\tilde d}^2$ is written as

\begin{equation}
M_{\tilde d}^2 = \pmatrix{~m_{0L}^2 I + \hat M_d^2
+c K^\dagger \hat
 M_u^2 K~ & ~ Am_{0}\hat M_d  \cr ~Am_{0}\hat M_d & ~
 m_{0R}^2 I +\hat M_d^2},
\label{matr}
\end{equation}
in a basis where the $3\times 3$ down-type quark mass matrix is
diagonal.
The matrices $\hat M_u$ and $\hat
M_d$ are diagonal up- and down-type quark mass matrices
respectively and
$K$ is the Cabibbo-Kobayashi-Maskawa mixing matrix. For simplicity we
take $m_{0L}=m_{0R}$ and equal to the universal scalar mass $m_0$. In
order
to simplify the process of analytic diagonalization of
$M_{\tilde d}^2$ we
take the trilinear scalar coupling $A$ equal to zero, which does not
affect the result significantly. The $c$-parameter,
which is responsible for flavor-violating interactions, plays an
important
role in our numerical estimates of the $b\rightarrow s \widetilde g
\widetilde g$
branching ratio.
Some authors \cite{indians} take $c$ of order 0.01 or even lower,
while others
\cite{bhara}
suggest that $c$ can be somewhat larger in magnitude. As regards
the sign of
the $c$-parameter, $c<0$ is preferred in the MSSM. In this paper, we
treat $c$
as a phenomenological parameter to be experimentally constrained.

In the case $A=0$, $M_{\tilde d}^2$ is block-diagonal and only the
upper-left block needs to be diagonalized. The upper-left block can
be
written in the form
\begin{equation}
M_{\tilde d(3 \times 3)}^{2}=m_0^2\left [I+ \pmatrix{0 & 0 & 0 \cr
0 & 0 & 0 \cr
0 & 0 & b}
+c'\pmatrix{0 & 0 & 0 \cr 0 & 0 & \epsilon \cr
0 & \epsilon & 1} \right ],
\end{equation}
where we have neglected the masses of the $d$ and $s$ quarks with
respect to the mass
of the $b$ quark, and similarly $m_u$ and $m_c$ with respect to
$m_t$. The
modified parameter $c'$ is equal to $c {m^2_t}/{m^2_0}$ and $b$ is
${m^2_b}/{m^2_0}$. Only the two
leading terms were kept in the product $K^{\dagger} \hat M_u^2 K$,
namely those
proportional to $|K_{tb}|^2$ and $K^{\ast}_{ts}K_{tb}$, the first
being
taken equal to unity and the latter being
denoted by
$\epsilon$. The matrix that diagonalizes $M_{\tilde d(3 \times 3)}^2$
is found to be
\widetext
\begin{equation}
\widetilde U_{(3 \times 3)}={1 \over (2f)^{1/2}}
\pmatrix{(2f)^{1/2} & 0 & 0 \cr
0 & (f+b+c')^{1/2} & (f-b-c')^{1/2} \cr
0 & -2\epsilon c'(f+b+c')^{-1/2} & 2\epsilon c'(f-b-c')^{-1/2} },
\end{equation}
\narrowtext
where $f$
is a function of the variables $b$, $c'$ and $\epsilon$ defined by
\begin{equation}
f\left( b, c', \epsilon \right )=\sqrt{\left (b+
c' \right )^2+4\epsilon^2c'^2}.
\end{equation}
The complete diagonalizing matrix $\widetilde U$ is given by
\begin{equation}
\widetilde U=\pmatrix{
\widetilde U_{(3 \times 3)} & 0 \cr
0 & I },
\end{equation}
where $I$ is the $(3 \times 3)$ identity matrix.
The evaluation of $\widetilde U^{\dagger}M_{\tilde d}^2
\widetilde U$ gives a diagonal matrix with the squark masses squared
on the diagonal.
For the left handed mass-eigenstates one gets
\begin{eqnarray}
m^2_{\tilde d_L} &=& m^2_0, \\
m^2_{\tilde s_L} &=& m^2_0\left (1 +{b \over 2} +{c'\over 2} -
{1 \over 2}f \left ( b, c', \epsilon  \right )\right ), \\
m^2_{\tilde b_L} &=& m^2_0\left(1 +{b \over 2} +{c'\over 2} +
{1 \over 2}f \left ( b, c', \epsilon \right )\right ),
\label{masbsquark}
\end{eqnarray}
while the right handed ones get masses
\begin{equation}
m^2_{\tilde d_R} = m^2_0+m_d^2,~~
m^2_{\tilde s_R} = m^2_0 +m^2_s,~~
m^2_{\tilde b_R} = m^2_0 +m^2_b.
\end{equation}

The matrix $\Gamma_L$ needed for calculation of the decay rate
$\Gamma (b \rightarrow s \tilde g \tilde g)$
can be written in the following way
\begin{equation}
\Gamma_L=\widetilde U^{\dagger}\pmatrix{I \cr 0}=
\pmatrix{\widetilde U_{(3 \times 3)}^{\dagger} \cr 0 }.
\end{equation}
Note that  the matrix $\widetilde
U_{(3 \times 3)}^{\dagger}$ reduces to
the identity matrix in the limit $c\rightarrow 0$,
as it should.
The matrix $\Gamma_R$ is trivially found to be
\begin{equation}
\Gamma_R=\pmatrix{~0 \cr ~I}.
\end{equation}

Having found the exact form of the matrices $\Gamma_L$ and
$\Gamma_R$,
the calculation of $\Gamma(b\rightarrow s \widetilde g \widetilde g
)$ can be completed analytically. The invariant matrix element
$\cal M$
consists of terms corresponding
to the exchange of $\widetilde b_L, \widetilde b_R, \widetilde s_L
\text{ and }
\widetilde s_R$. We have neglected much smaller terms with
$\widetilde d_L$ or
$\widetilde d_R$ exchange.
After performing the 3-body Lorentz invariant phase space
integration,
the decay rate becomes
\widetext
\begin{equation}
\Gamma (b \rightarrow s \tilde g \tilde g)=2{\alpha_S^2m^5_b
\over 54 \pi}I \left ( {m_s \over m_b} \right )
\sum_{i,j}{1 \over m^2_im^2_j}
\left (
\Gamma^{ib}_L
\Gamma^{\dagger bj}_L+ \Gamma^{ib}_R \Gamma^{\dagger bj}_R \right )
\left(
\Gamma^{\dagger si}_L\Gamma^{js}_L+ \Gamma^{\dagger si}_R
\Gamma^{js}_R \right ),
\label{firstdr}
\end{equation}
\narrowtext
for $i,j=\widetilde b_L, \widetilde b_R, \widetilde s_L, \widetilde
s_R$.
The function $I(x)$ is given by
\begin{equation}
I(x)=1-8x^2+24x^4\ln x+8x^6-x^8.
\end{equation}
The overall multiplicative factor of 2 in Eq.\ (\ref{firstdr}) is
due to the Majorana nature of
the external
gluinos. Using the ``diagonal'' character of $\Gamma_R$ this can be
futher reduced
to
\begin{equation}
\Gamma(b \rightarrow s \tilde g \tilde g)={\alpha_S^2m^5_b \over 27
\pi}I \left ( {m_s \over m_b} \right )
\sum_{i,j}{1 \over m^2_im^2_j}
\Gamma^{ib}_L
\Gamma^{\dagger bj}_L\Gamma^{\dagger si}_L\Gamma^{js}_L.
\end{equation}
The sum can be written
in terms of the squark masses
\begin{equation}
\sum_{i,j}{1 \over m^2_im^2_j}
\Gamma^{ib}_L
\Gamma^{\dagger bj}_L\Gamma^{\dagger si}_L\Gamma^{js}_L={\epsilon ^2
c'^2
\over f^2}{\left (m_{\tilde b_L}^2 - m_{\tilde s_L}^2 \right )^2
\over
m_{\tilde b_L}^4m_{\tilde s_L}^4}.
\end{equation}
Using the expressions for the squark masses obtained as the
eigenvalues of $M_{\tilde d}^2$
we can write the result for $\Gamma(b\rightarrow s \tilde g
\tilde g)$ in the form
\begin{eqnarray}
\Gamma(b\rightarrow s \tilde g \tilde g)=&&
{\alpha_S^2m^5_b \over 27 \pi}I \left ( {m_s \over m_b}
\right ) {m^4_t \over
m^8_0}|K^\ast_{ts}|^2 \nonumber \\
&&\times \left ({c \over 1+{m^2_b\over m^2_0}+c{m^2_t
\over m^2_0}-c^2|K^\ast_{ts}|^2{m^4_t \over m^4_0}}\right )^2.
\label{lastdr}
\end{eqnarray}
Both the terms $m^2_b/m^2_0$ and $c^2|K^\ast_{ts}|^2{m^4_t \over
m^4_0}$ in
the denominator can
be neglected with respect to $c{m^2_t\over m^2_0}$, if $m_0$ is
larger than
80 GeV. Note also that $\Gamma(b \rightarrow s \tilde g \tilde g)$
cannot
develop a pole because of the experimental lower limit on the masses
of
squarks in the intermediate state.
For example, according to \cite{ccfhpy} we can require in the light
gluino
case that $m_{\tilde b_L}\geq 60 \text{GeV}$. As will be
discussed below, this imposes an
additional constraint on the $c$ parameter as a function of $m_0$.

It is convenient to define the ratio
\begin{equation}
R_{s\tilde g \tilde g}={\Gamma (b \rightarrow s \tilde g \tilde g)
\over
\Gamma (b \rightarrow c \bar u d)+\Gamma (b \rightarrow c \bar u s)},
\label{defr}
\end{equation}
where the denominator is given by
\begin{eqnarray}
\Gamma(b\rightarrow c\bar ud)+&&\Gamma(b\rightarrow c\bar us)
\nonumber\\
&&={3G_F^2m^5_b|K_{cb}|^2 \over 192 \pi^3}
I_0 \left ({m^2_c \over m^2_b},0,0 \right )\eta J.
\label{defdenr}
\end{eqnarray}
The expression for the phase space factor $I_0$ and the values of
leading-log anomalous
dimension enhancement $\eta$ and next-to-leading corrections
enhancement $J$
can be found in the literature (e.g. \cite{bigi,kagan}) and their
product is of order
$O(1)$.
Combining equations (\ref{lastdr}), (\ref{defr}) and (\ref{defdenr})
we get
\begin{equation}
R_{s\tilde g \tilde g}={128 \over 27}\left ( {\alpha_S \over
\alpha} \right )^2 {\sin^4(\theta_W)c^2\over \left (m_0^2+cm_t^2
\right )^2}\left ({m_Wm_t \over m_0} \right )^4\bigg |{K^\ast_{ts}
\over K_{cb}}
\bigg |^2.
\label{resultr}
\end{equation}

If the ratio $R_{s\tilde g \tilde g}$ gets as high as 20\%, then the
branching ratio of the non-standard model decay $b \rightarrow s
\tilde g
\tilde g$ is more than 14\%. This could completely account for the
discrepancy
between $BR_{SL}(B) \Bigl | _{exp}$
and $BR_{SL}(B) \Bigl | _{QCD}$. The current experimental data can
be fit if
$R_{s\tilde g \tilde g}=(22.68\pm 0.52)\%$.

In general, by requiring $R_{s \tilde g \tilde g}$ to have a certain
value,
one gets
$c$ as a function of the universal scalar mass $m_0$. This function
is plotted in Fig. 1 for several different values of $R_{s\tilde g
\tilde g}$.
We have used $\alpha_S(m_b)=0.18$, $\alpha(m_b)=1/133$, $\sin ^2
\theta_W=0.232$
and $m_t=170 \text{GeV}$ in all numerical calculations. Also, we
have taken
advantage of the equality $|K_{ts}|\simeq |K_{cb}|$.
{}From the first figure it can be seen that the needed nonleptonic
enhancement in $b$ decays
can be obtained using the contribution from the process $b
\rightarrow s \tilde g
\tilde g$ for reasonable values of $m_0$ and $c$. The necessary
values of $c$
as a function of $m_0$ are intermediate between those considered
by refs
\cite{indians} and \cite{hagelin}.

The lower bound on the $b$ squark mass of 60 GeV mentioned above
does not
interfere with any of the curves plotted in Fig. 1. In fact, the
mass of the $b$
squark is certainly above 75 GeV for $m_0\geq 80 \text{GeV}$. The
$m_{\tilde b_L}$
as a function of $m_0$ obtained using equations (\ref{masbsquark})
and
(\ref{resultr}) is plotted in Fig. 2.
This has an interesting implication for the problem of the $b$
excess in
$Z$ decays.
A possible explanation of the
$b$ anomaly could have been the $Z$ decay into
the $b$ squark and the $b$ anti-squark. For that to be possible,
the $b$ squarks would have
to have a mass less than $M_Z/2$. In order to explain the
$b$-anomaly using
this process, one would need the ratio $\Gamma (Z\rightarrow \tilde b
\bar {\tilde b})/\Gamma(Z\rightarrow b \bar b)$ to be about 2\%. This
is possible,
however, only if $m_{\tilde b_L}\approx 0.47M_Z$. This imposes
the following constraint on the $c$-parameter through Eq.\
(\ref{masbsquark})
\begin{equation}
c={\left (0.47M_Z\right )^2-m^2_0 \over m^2_t}.
\end{equation}
Unfortunatelly, this
constraint is incompatible with the $c$ dependence on $m_0$ that we
got from the
analysis of the $B$ semileptonic branching ratio. In addition such
a light
$b$ squark would lead to unacceptably large contributions from $Z
\rightarrow \bar b
\tilde b \tilde g$.
Therefore, in the light gluino scenario one has to
rule out the possibility of $b$ squarks being lighter than $M_Z/2$.
Nevertheless, there are other mechanisms, that can
explain the $b$ excess \cite{clavelli} without contradiction with our
current analysis.
The current calculation is not sensitive to the gluino mass varying
in the
range of the low mass window (0 GeV to 0.7 GeV).

It is interesting to compare the decay rate (\ref{lastdr}) to the
decay rate of $b\rightarrow s g$,
because processes like this could also account
for the missing 14-20\% in the hadronic branching ratio of the $B$
\cite{kagan}.
We use the formula for $\Gamma_{SUSY}(b\rightarrow s g)$ given in
\cite{bertolini},
that corresponds to the processes with a squark and a gluino
exchange within a loop and
an external gluon attached either to the gluino line or the squark
line. The decay rate
is given by
\begin{eqnarray}
\Gamma_{SUSY}(b\rightarrow s g)=&&{\alpha_S^3 \over 16 \pi^2}m^5_b
\left (
1- {m^2_s \over m^2_b} \right)^3 \left ( 1+{m^2_s \over m^2_b}
\right )
\nonumber\\&&\times
\left ({cm^2_t \over \tilde m^4}\right )^2|K_{tb}K_{ts}^{\ast}|^2
\nonumber\\&&\times
\bigg \{A \sqrt x\left [ {1 \over 3} g_d(x)-3g_c(x) \right ]
\nonumber\\&&
\phantom{\times \bigg \{}
- \left [
{1 \over 6} f_b(x)-{3 \over 2}f_a(x) \right ] \bigg \}^2,
\end {eqnarray}
with $x={m^2_{\tilde g}}/{\tilde m^2}$ and $\tilde m^2={1 \over 2}
\left (m^2_{\tilde b}+m^2_{\tilde s} \right )$. The functions $g$
and $f$ are
given in \cite{bertolini} but for the purpose of our comparison we
need only their
limits as $x\rightarrow 0$. These are
\begin{equation}
\lim_{x\rightarrow 0}f_a(x)={1 \over 3};\quad  \lim_{x\rightarrow
0}f_b(x)={1
\over 6}.
\end{equation}
The decay rate in the case of the gluino with a negligible mass is
then equal to
\begin{equation}
\widetilde \Gamma_{SUSY}(b\rightarrow s g)={289 \over 20736}
{\alpha_S^3 \over \pi^2}
m^5_b {c^2m^4_t \over \tilde m^8}|K_{tb}K_{ts}^{\ast}|^2,
\label{drbsg}
\end {equation}
where the terms proportional to ${m^2_s}/{m^2_b}$ and its higher
powers were neglected.
Dividing Eq.\ (\ref{drbsg}) by Eq.\ (\ref{lastdr}) one gets
\begin{equation}
{\widetilde \Gamma_{SUSY}(b\rightarrow s g) \over \Gamma(b\rightarrow
s \tilde g
\tilde g)}={289 \over 768}{\alpha_S \over \pi},
\end{equation}
indicating that the contribution to the total $b$ width from
$b\rightarrow
s \tilde g \tilde g$ decay is dominant over the one from $b
\rightarrow s g$
in the light gluino scenario.
The mechanism of \cite{bertolini} for the non-leptonic enhancement
is only consistent with a $b$
squark above $M_Z/2$ if the gluino is heavier than 2 GeV
and $m_0$ is less than 150 GeV.

In conclusion we can say that, assuming the gluino is light, the
decay
$b\rightarrow s \tilde g \tilde g$ provides a plausible explanation
of the gap
between $BR_{SL}(B) \Bigl | _{exp}$ and $BR_{SL}(B)
\Bigl | _{QCD}$. The
final state gluinos in $b$ decay could hadronize into the
gluino-gluon or
gluino-gluino bound
states discussed in \cite{farrar} or merely into intrinsic gluino
components of normal hadrons. Since
the values of the $c$-parameter and the universal scalar mass are
not yet
well determined, the branching ratio of $b\rightarrow s \tilde g
\tilde g$
may provide a useful constraint as experiments improve.

We acknowledge useful and stimulating discussions with P. Cox.
This work
was supported in part by the U.S. Department of Energy under Grant
No.
DE-FG05-84ER40141.

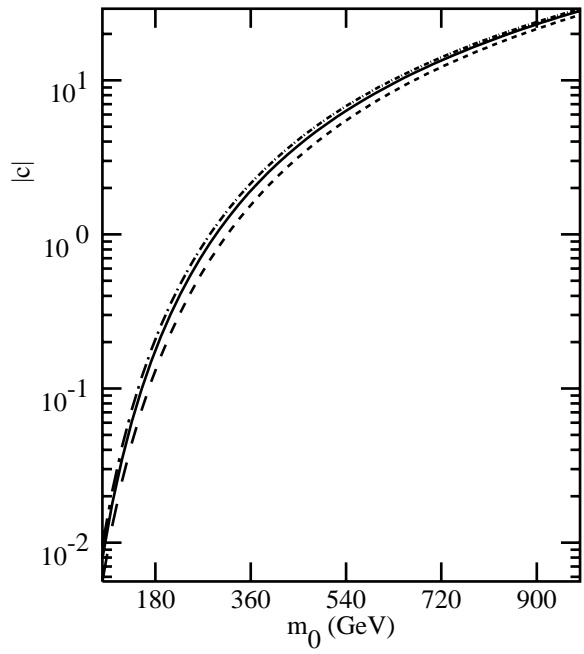
\begin{figure}
\caption{
The absolute value of $c$ ($c$ is assumed to be negative) is plotted
as a function of the universal scalar mass $m_0$ for $R_{s\tilde g
\tilde g}=$
0.3 (dot-dashed line), 0.2 (solid line) and 0.1 (dashed line).
The mass of the $t$ quark was taken to be 170 GeV.}
\end{figure}

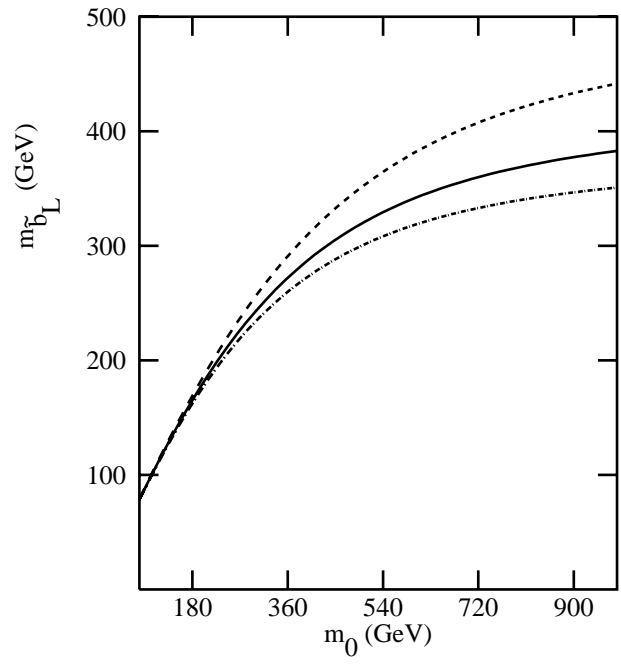
\begin{figure}
\caption{
The mass of the $b_L$ squark is plotted as a function of the
universal
scalar mass $m_0$ for $R_{s\tilde g \tilde g}=$
0.3 (dot-dashed line), 0.2 (solid line) and 0.1 (dashed line).
The mass of the $t$ quark was taken to be 170 GeV.}
\end{figure}

\end{document}